# The influence of boundary conditions and interfacial slip on the time taken to achieve a nonequilibrium steady-state for highly confined flows

Carmelo Riccardo Civello[1], Luca Maffioli[1], Joseph Johnson[2],

Edward R. Smith[3], James P. Ewen[4], Peter J. Daivis[5],

Daniele Dini[4], B. D. Todd[1]

[1]Department of Mathematics, Faculty of Science, Computing and Emerging Technologies, Swinburne University of Technology, PO Box 218, Hawthorn, Victoria 3122, Australia

[2]School of Mathematics and Statistics, University of Melbourne, Victoria 3010, Australia

[3]Mechanical and Aerospace Engineering, Brunel University London, Kingston Lane, Uxbridge, UB8 3PH, London, United Kingdom

[4]Department of Mechanical Engineering, Imperial College London, South Kensington Campus, London, SW7 2AZ, London, United Kingdom

[5]School of Science, RMIT University, GPO Box 2476, Melbourne, 3001, Victoria, Australia

Corresponding author: B. D. Todd, `btodd@swin.edu.au`

September 25, 2025

## Abstract

In this work we investigate the equilibration time to attain steady-state for a system of liquid molecules under boundary-driven planar Couette flow via nonequilibrium molecular dynamics (NEMD) simulation. In particular, we examine the equilibration time for the two common types of boundary driven flow: one in which both walls slide with equal and opposite velocity ($\pm \hat{u}/2$), and the other in which one wall is fixed and the other moves with twice the velocity ($\hat{u}$). Both flows give identical steady-state strain rates, and hence flow properties, but the transient behaviour is completely different. We find that in the case of no-slip boundary conditions, the equilibration times for the counter-sliding walls flow are exactly 4 times faster than those of the single sliding wall system, and this is independent of the atomistic nature of the fluid, i.e., it is an entirely hydrodynamic feature. We also find that systems that exhibit slip have longer equilibration times in general and the ratio of equilibration times for the two types of boundary-driven flow is even more pronounced. We analyse the problem by decomposing a generic planar Couette flow into a linear sum of purely symmetric and antisymmetric flows. We find that the no-slip equilibration time is dominated by the slowest decaying eigenvalue of the solution to the Navier-Stokes equation. In the case of slip, the longest relaxation time is now dominated by the transient slip velocity response, which is longer than the no-slip response time. In the case of a high-slip system of water confined to graphene channels, the enhancement is over two orders of magnitude. We propose a simple universal relation that predicts the enhanced equilibration time, which agrees well with our NEMD results for simple Lennard-Jones fluids and the water-graphene system. The implications of this significant speed-up in attaining steady-state, which is especially pronounced in the presence of slip, is discussed in general.

# 1  Introduction

Nonequilibrium molecular dynamics (NEMD) simulations have proven to be a valuable complement to experimental techniques in the study of liquid slip on various solid surfaces [1–3]. However, NEMD is constrained by its inherent length and time scales and associated statistical uncertainty [4–7]. Therefore, striking a balance between accuracy and computational cost is crucial. In light of these constraints, this study demonstrates that in the case of boundary-driven flows, selecting appropriate boundary conditions (BCs) can accelerate the transition to a nonequilibrium steady-state, vastly improving computational efficiency whilst maintaining accuracy. This phenomenon will be seen to be even more significant for systems where the Navier no-slip boundary condition is not met [8]. Slip is extremely important in various applications, such as nanofluidics [9], slippery liquid-infused porous surfaces [10] and tribology [11].

It is well recognized that a direct application of NEMD achieves good signal-to-noise ratios (SNR) only at shear rates orders of magnitude higher than those characterizing real experiments or commonly experienced physical processess [12, 13]. At low shear rates, thermal fluctuations dominate the signal, necessitating computationally infeasible averaging which renders the NEMD method ineffective. To address this issue, a multiscale method was recently proposed for studying polymer melts under shear flow [14]. The multiscale approach reduces the computational cost of simulations while extending the range of accessible shear rates. However, if one is only concerned with capturing the full nonlinear response of a molecularly detailed fluid, then the only proven technique allowing the practitioner to sample physically realisable strain rates is the Transient Time Correlation Function (TTCF) method [5, 7, 12, 13, 15–18]. TTCF directly links a system's equilibrium state to its nonequilibrium steady-state, and its effectiveness strongly depends on computing the transient time evolution of trajectories in phase-space. On the other hand, the signal-to-noise ratio (SNR) of TTCF results decreases linearly over time [13]. Hence, when investigating steady states, the transient response time impacts the computational cost of simulations, which is a general concern in molecular dynamics simulations. In this context, we will show that selecting appropriate boundary conditions is crucial, as it allows the system to reach a steady state through a significantly shorter transient response time.

Over the past few decades, researchers have relied on NEMD simulations to study boundary slip at the liquid-solid interface. Significant progress has been made in understanding its dependence on factors such as temperature [19–22], surface wettability [23, 24], surface roughness [25–27], pressure [28, 29] and shear rate [14, 30–32]. However, a unified model is still lacking. Boundary slip has been extensively studied due to its impact on system properties, including flow enhancement and friction reduction [11, 33–36]. For example, carbon nanotubes (CNTs) have attracted significant attention due to their exceptional properties. Among these are a high surface-to-volume ratio and high water transport efficiency, facilitated by their hydrophobicity and surface smoothness. In fact, water exhibits exceptionally high slippage at the interface with smooth carbon surfaces [36–39]. In what follows, we will demonstrate that the choice of boundary conditions has an even greater impact on the time evolution of this type of system when interfacial slip is taken into account, and that this impact varies with the degree of slip.

When it comes to boundary driven NEMD, no specific attention has been placed on which boundary conditions to employ. There are studies where only one of the two surfaces is moving at constant velocity to shear the confined fluid, while the other is kept fixed [22, 24, 30, 38, 40–42], as well as studies where both surfaces are moving at the same velocity, but in opposite directions [12, 21, 28, 43]. Given the increasing use of NEMD simulations, the relevance of computational cost optimization and the lack of a shared criterion motivates our focus on how the boundary conditions affect the time-evolution of fluid systems in general. A direct corollary leads us to investigate the process when interfacial slip is considered. We will find that although the steady-states of both types of boundary driven flows are identical, as of course they must be, the transient behaviours are not and this has a significant impact on the equilibration times.

In what follows, the physics of time evolving boundary driven planar Couette flow systems is discussed in detail, using continuum theory. NEMD results for an atomistic Lennard-Jones fluid are fitted to the continuum models and excellent agreement will be demonstrated for sufficiently wide channels. Results from a simulated high-slip water-graphene system are also included to highlight the effect that slip has on the equilibration rate. With this study we show that there are preferential BCs that reduce the computational cost of boundary-driven simulations by at least a factor of four, and this factor substantially increases when interfacial slip is introduced. In the case of water confined by graphene walls, the improvement is several orders of magnitude. We propose a universal relationship that predicts the speed up as a function of the degree of slip. The implications in general for diffusive transport processes and tribology related technologies are also briefly discussed in the conclusions.

## 2 Theoretical discussion

This section presents the derivation of the analytical solution for planar Couette flow under three different boundary conditions: no slip, slip with no time dependence, and time dependent slip. Particular attention is given to the symmetry of the flow and its influence on the transient evolution towards steady-state. Full details of the derivations are provided in the Supplemental Information accompanying this paper.

### 2.1 No-slip boundary conditions

We study a fluid confined between two parallel walls separated by a distance $L$ in the $z$-direction and subjected to shear in the $x$-direction. Assuming laminar incompressible flow, the related Navier-Stokes momentum equation is

$$\frac{\partial u}{\partial t} = \nu \frac{\partial^2 u}{\partial z^2}, \qquad (1)$$

where $u = u_x(z,t)$ is the $x$ component of the fluid velocity as a function of the position $z$ across the channel and $\nu$ is the kinematic viscosity of the fluid. For a generic Couette flow, we assume the upper (top) and lower (bottom) walls move at velocities $u_T$ and $u_B$ in the $x$-direction, respectively, and the fluid is at rest at $t = 0$. The related Cauchy problem is

$$\begin{cases} \dfrac{\partial u}{\partial t} = \nu \dfrac{\partial^2 u}{\partial z^2} & z \in [-L/2, L/2] \text{ and } t > 0, \\ u(L/2, t) = u_T & t > 0, \\ u(-L/2, t) = u_B & t > 0, \\ u(z, 0) = 0 & z \in [-L/2, L/2]. \end{cases} \qquad (2)$$

The general solution can be found using the trigonometric representation:

$$\begin{aligned} u(z,t) = &\frac{u_T + u_B}{2} + \frac{u_T - u_B}{L} z \\ &+ \sum_{n=0}^{\infty} \Bigg\{ (u_T + u_B) \frac{2(-1)^{n+1}}{(2n+1)\pi} \exp\left\{ -\nu \left[\frac{(2n+1)\pi}{L}\right]^2 t \right\} \cos\left[\frac{(2n+1)\pi}{L} z\right] \\ &+ (u_T - u_B) \frac{(-1)^{n+1}}{(n+1)\pi} \exp\left\{ -\nu \left[\frac{(2n+2)\pi}{L}\right]^2 t \right\} \sin\left[\frac{(2n+2)\pi}{L} z\right] \Bigg\}. \end{aligned} \qquad (3)$$

Depending on the values of the wall velocities, one can have a combination of cosine (even) and sine (odd) functions. However, the former are related solely to the *symmetric* part of the boundary conditions, $u_T + u_B$, while the latter depends on the *antisymmetric* component, $u_T - u_B$. By the following change of variables

$$\bar{u}_s \equiv \frac{u_T + u_B}{2} \text{ and } \bar{u}_a \equiv \frac{u_T - u_B}{2} \longrightarrow u_B = \bar{u}_s - \bar{u}_a \text{ and } u_T = \bar{u}_s + \bar{u}_a, \qquad (4)$$

the flow can be split into symmetric and antisymmetric components, with the related Cauchy problems

$$u(z,t) = u_s(z,t) + u_a(z,t), \qquad (5)$$

$$\begin{cases} \dfrac{\partial u_s}{\partial t} = \nu \dfrac{\partial^2 u_s}{\partial z^2}, \\ u_s(-L/2,t) = \bar{u}_s, \\ u_s(L/2,t) = \bar{u}_s, \\ u_s(z,0) = 0, \end{cases} \qquad \begin{cases} \dfrac{\partial u_a}{\partial t} = \nu \dfrac{\partial^2 u_a}{\partial z^2}, \\ u_a(-L/2,t) = -\bar{u}_a, \\ u_a(L/2,t) = \bar{u}_a, \\ u_a(z,0) = 0, \end{cases} \qquad (6)$$

which have the following solutions

$$u_s(z,t) = \bar{u}_s + \sum_{n=0}^{\infty} \frac{4\bar{u}_s(-1)^{n+1}}{\lambda_{s,n}^{(0)} L} \exp[-\nu(\lambda_{s,n}^{(0)})^2 t] \cos(\lambda_{s,n}^{(0)} z),$$

$$u_a(z,t) = \frac{2\bar{u}_a}{L} z + \sum_{n=0}^{\infty} \frac{4\bar{u}_a(-1)^{n+1}}{\lambda_{a,n}^{(0)} L} \exp[-\nu(\lambda_{a,n}^{(0)})^2 t] \sin(\lambda_{a,n}^{(0)} z), \qquad (7)$$

$$\text{where } \lambda_{s,n}^{(0)} = \frac{(2n+1)\pi}{L} \text{ and } \lambda_{a,n}^{(0)} = \frac{(2n+2)\pi}{L}.$$

The linearity of the underlying governing equations allows us to separately examine the symmetric and antisymmetric flows. This linearity persists even in the presence of slip enabling us to continue to examine these flows separately despite the fact that the trigonometric functions are not orthogonal in the presence of slip. Moreover, the symmetric part of the flow is associated with odd wave numbers $\lambda_{s,n}^{(0)} = (2n+1)\pi/L$, while the antisymmetric part is characterized by even wave numbers $\lambda_{a,n}^{(0)} = (2n+2)\pi/L$ and the superscript '(0)' denotes the absence of slip. The wave numbers also determine the typical time scale of the flow and the speed of convergence to the steady-state is determined by the slowest mode, i.e. the first wave numbers. Since

$$\left( \frac{\lambda_{s,0}^{(0)}}{\lambda_{a,0}^{(0)}} \right)^2 = \frac{1}{4}, \qquad (8)$$

the antisymmetric component of the flow converges exactly four times faster than the symmetric part. As the actual velocity profile is a linear combination of the symmetric and antisymmetric flows, the slowest convergence time is given by the symmetric wave number, i.e. $\lambda_{s,0}^{(0)}$. This is a key observation of this analysis. We further emphasise that this ratio is independent of the molecular details of the fluids, i.e., it is purely a hydrodynamic feature. Physically, it means that a fluid confined by a channel of width $L$ and driven by confining walls moving at equal velocity magnitude $\hat{u}/2$ but opposite directions (a purely antisymmetric flow) will reach steady-state four times faster than the same fluid driven by moving just one wall at velocity $\hat{u}$, whilst keeping the other wall fixed (a combination of symmetric and antisymmetric flows). The physical mechanism for this is of course momentum diffusion, which occurs on a timescale of $L^2/(4\pi^2 \nu)$ for the former, compared to $L^2/(\pi^2 \nu)$ for the latter. Finally, it should be noted that this is independent of the absolute value of the wall velocity and that the actual convergence times scale as $L^2$ for both flows.

## 2.2 Slip boundary conditions with no time dependence

In this section, we assume the system has velocity slip at the walls but that it is constant for all times, $t > 0$. For the counter-sliding wall geometry depicted by Fig. 1(a), the fluid velocity consists of purely antisymmetric components:

$$u_T \equiv \frac{\hat{u}}{2} \text{ and } u_B = -\frac{\hat{u}}{2} \longrightarrow \bar{u}_s = 0 \text{ and } \bar{u}_a = \frac{\hat{u}}{2}, \tag{9}$$

whereas a flow with equivalent steady-state shear rate, generated by moving the top wall at velocity $\hat{u}$ and keeping the bottom wall fixed (Fig. 1(b)) is composed of a linear combination of symmetric and antisymmetric components:

$$u_T \equiv \hat{u} \text{ and } u_B = 0 \longrightarrow \bar{u}_s = \frac{\hat{u}}{2} \text{ and } \bar{u}_a = \frac{\hat{u}}{2}, \tag{10}$$

and hence is expected to converge to a steady state more slowly as discussed at the end of the previous section.

This effect is, as we will show, enhanced by the presence of slip. The simplest model that includes slip is the Navier slip hypothesis, which imposes a linear relation between slip velocity and shear pressure

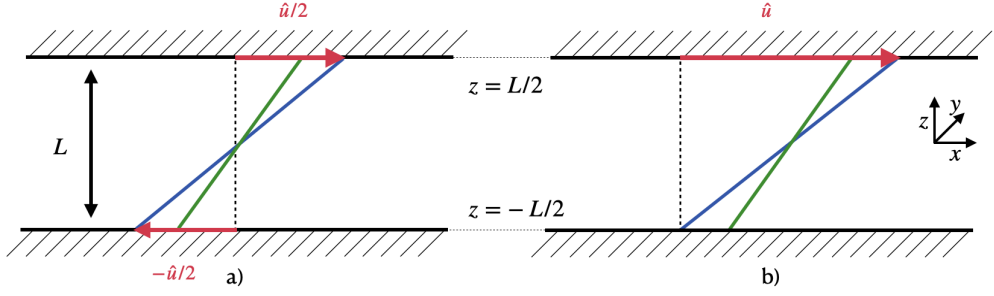

**Figure 1:** (a) Counter-sliding (antisymmetric) and (b) single sliding-wall setups. The blue lines describe the no-slip steady-state, whereas the green lines depict a generic steady-state with slip.

at the wall-fluid interface:

$$\frac{\partial u}{\partial z}\left(-\frac{L}{2},t\right) = -\frac{1}{L_s}\left[u_B - u\left(-\frac{L}{2},t\right)\right] \text{ and } \frac{\partial u}{\partial z}\left(\frac{L}{2},t\right) = \frac{1}{L_s}\left[u_T - u\left(\frac{L}{2},t\right)\right], \quad (11)$$

where $L_s$ is the slip length associated with the system. By applying the same change of variables as before we have

$$\begin{cases} \frac{\partial u_s}{\partial t} = \nu \frac{\partial^2 u_s}{\partial z^2}, \\ \frac{\partial u_s}{\partial z}\left(-\frac{L}{2},t\right) = -\frac{1}{L_s}\left[\bar{u}_s - u_s\left(-\frac{L}{2},t\right)\right], \\ \frac{\partial u_s}{\partial z}\left(\frac{L}{2},t\right) = \frac{1}{L_s}\left[\bar{u}_s - u_s\left(\frac{L}{2},t\right)\right], \\ u_s(z,0) = 0, \end{cases} \quad \begin{cases} \frac{\partial u_a}{\partial t} = \nu \frac{\partial^2 u_a}{\partial z^2}, \\ \frac{\partial u_a}{\partial z}\left(-\frac{L}{2},t\right) = -\frac{1}{L_s}\left[-\bar{u}_a - u_a\left(-\frac{L}{2},t\right)\right], \\ \frac{\partial u_a}{\partial z}\left(\frac{L}{2},t\right) = \frac{1}{L_s}\left[\bar{u}_a - u_a\left(\frac{L}{2},t\right)\right], \\ u_a(z,0) = 0, \end{cases} \quad (12)$$

and finally

$$\begin{aligned} u_s(z,t) &= \bar{u}_s + \sum_{n=0}^{\infty} C_{s,n}(t) \cos(\lambda_{s,n} z), \\ u_a(z,t) &= \frac{2\bar{u}_a}{L+2L_s} z + \sum_{n=0}^{\infty} C_{a,n}(t) \sin(\lambda_{a,n} z). \end{aligned} \quad (13)$$

The Fourier coefficients $C_{s,n}$ and $C_{a,n}$ are

$$\begin{aligned} C_{s,n}(t) &= -\frac{4\bar{u}_s \sin(L\lambda_{s,n}/2)}{L\lambda_{s,n} + \sin(L\lambda_{s,n})} \exp(-\nu \lambda_{s,n}^2 t), \text{ and} \\ C_{a,n}(t) &= \frac{4\bar{u}_a \left[L\lambda_{a,n} \cos(L\lambda_{a,n}/2) - 2\sin(L\lambda_{a,n}/2)\right]}{(L+2L_s)\lambda_{a,n} \left[L\lambda_{a,n} - \sin(L\lambda_{a,n})\right]} \exp(-\nu \lambda_{a,n}^2 t), \end{aligned} \quad (14)$$

while the wave numbers are determined by the following nondimensionalised transcendental relations derived from the boundary conditions:

$$\tan\left(\tilde{\lambda}_{s,n}\right) = \frac{1}{\tilde{L}_s \tilde{\lambda}_{s,n}} \text{ and } \cot\left(\tilde{\lambda}_{a,n}\right) = -\frac{1}{\tilde{L}_s \tilde{\lambda}_{a,n}}, \tag{15}$$

where $\tilde{\lambda} \equiv \frac{\lambda L}{2}$ and $\tilde{L}_s \equiv \frac{2L_s}{L}$.

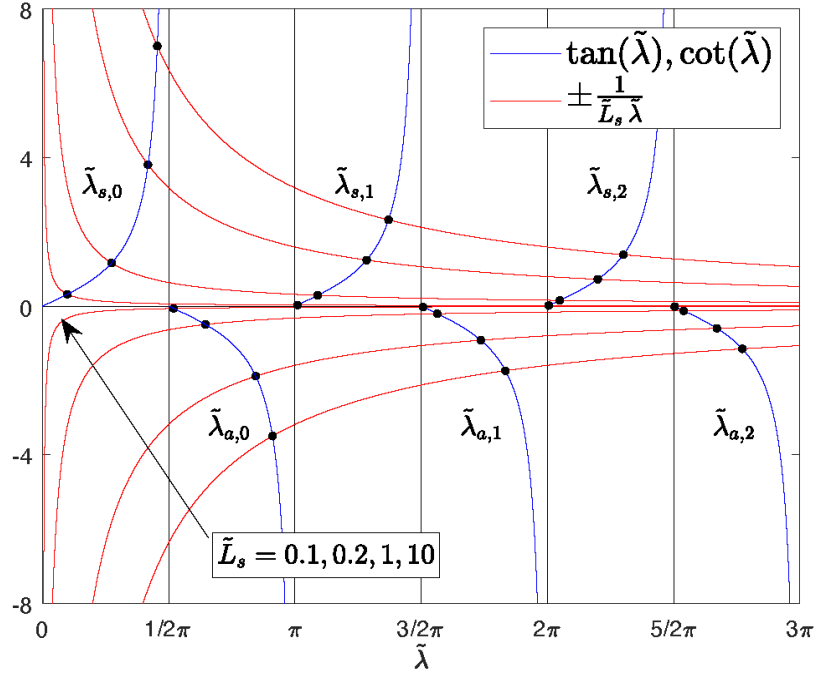

**Figure 2:** Tangent and cotangent of the wave numbers (blue lines, upper and lower half respectively), and right hand sides of Eqs (16) (red lines, upper half for the first equation, lower half for the second equation), as functions of different slip lengths. The wave numbers and the slip length have been nondimensionalized by the transformation $\tilde{\lambda} = \lambda L/2$ and $\tilde{L}_s = 2L_s/L$. The black dots are the intersection of the two functions, i.e., the actual wave number associated with the system. The first three symmetric and antisymmetric wave numbers are displayed here for reduced slip lengths $\tilde{L}_s$ of $0.1, 0.2, 1, 10$. The arrow indicates the direction of increasing slip length, and the same values have been used for both sets of curves. As the slip length increases, all wave numbers shift toward lower values, indicating an increasing transient.

Eq. (15) cannot be solved in a closed form; however, the wave numbers can be computed numerically as the roots of these equations, and they are shown graphically in Fig. 2. The quantitative behaviour can be summarized by the following relations

$$\begin{cases} \lim_{\tilde{L}_s \to \infty} \tilde{\lambda}_{s,n} = n\pi, \\ \lim_{\tilde{L}_s \to \infty} \tilde{\lambda}_{a,n} = \dfrac{(2n+1)\pi}{2}, \end{cases} \qquad \begin{cases} \lim_{\tilde{L}_s \to 0^+} \tilde{\lambda}_{s,n} = \dfrac{(2n+1)\pi}{2}, \\ \lim_{\tilde{L}_s \to 0^+} \tilde{\lambda}_{a,n} = (n+1)\pi, \end{cases} \qquad (16)$$

from which we can draw the following conclusions:

(i) The wave numbers of the symmetric flow are always larger that those of the antisymmetric flow, hence the transient of the former is longer than the latter in any condition. This implies that a counter-sliding wall system will always attain steady-state faster than a single-sliding wall system.

(ii) The presence of any slip makes the wave numbers decrease, hence making the convergence time to the steady state longer.

(iii) In the limit of $\tilde{L}_s \to 0^+$, the relations of the no-slip case are obtained, and finally

(iv) In the limit of weak wall-fluid interaction ($\tilde{L}_s \to \infty$) the transient of the symmetric flow becomes indefinitely long ($\tilde{\lambda}_{s,0} = 0$), while the antisymmetric flow converges in finite time under any condition ($\tilde{\lambda}_{a,0} \neq 0$). The implication for high-slip systems, such as water-graphene, is that convergence to the steady-state is significantly faster for a counter-sliding wall system compared to a single-sliding one.

## 2.3 Time-dependent boundary conditions

In this section we assume the walls move with constant speed, but the slip is now time-dependent, which we express as a time-dependent boundary condition. Eq. (2) can be generalised by allowing the boundary velocity to vary with time. In such a case, Eq. (2) can be re-written as

$$\begin{cases} \dfrac{\partial u}{\partial t} = \nu \dfrac{\partial^2 u}{\partial z^2} & z \in [-L/2, L/2] \text{ and } t > 0, \\ u(L/2, t) = u_T(t) & t > 0, \\ u(-L/2, t) = u_B(t) & t > 0, \\ u(z, 0) = 0 & z \in [-L/2, L/2]. \end{cases} \qquad (17)$$

The general solution is similarly found as

$$u(z,t) = \frac{u_T(t) + u_B(t)}{2} + \frac{u_T(t) - u_B(t)}{L} z + \sum_{n=0}^{\infty} \left[ S_{s,n}(t) \cos\left(\lambda_{s,n}^{(0)} z\right) + S_{a,n}(t) \sin\left(\lambda_{a,n}^{(0)} z\right) \right]. \qquad (18)$$

The wave numbers are the same as those of the no-slip solution (Eq. (3)), whilst the coefficients of the trigonometric representation, $S_{s,n}(t)$ and $S_{a,n}(t)$, are now time-dependent:

$$S_{s,n}(t) = \frac{2(-1)^{n+1}\exp[-\nu(\lambda_{s,n}^{(0)})^2 t]}{\lambda_{s,n}^{(0)} L}\left\{u_T(0) + u_B(0) + \int_0^t [u_T'(\tau) + u_B'(\tau)]\exp[\nu(\lambda_{s,n}^{(0)})^2\tau]d\tau\right\},$$
$$S_{a,n}(t) = \frac{2(-1)^{n+1}\exp[-\nu(\lambda_{a,n}^{(0)})^2 t]}{\lambda_{a,n}^{(0)} L}\left\{u_T(0) - u_B(0) + \int_0^t [u_T'(\tau) - u_B'(\tau)]\exp[\nu(\lambda_{a,n}^{(0)})^2\tau]d\tau\right\},$$
(19)

where the prime symbol ' ' ' denotes the time derivative of a quantity. For such a model, the convergence to a steady-state is strictly related to the boundary functions $u_B(t)$ and $u_T(t)$. The transient phase is now no longer dominated solely by the first wave numbers, but also depends on the transient behaviour of the boundary functions. We will discuss this further in the next section when comparing the continuum models with our NEMD simulation results.

## 3 Method

### 3.1 Nonequilibrium molecular dynamics simulations

We simulate two distinct systems, one governed by simple Lennard-Jones interactions, and another consisting of water molecules confined by graphene walls. To simulate an inhomogeneous atomistic system governed by Lennard-Jones interactions, we adopt the same model as described by Maffioli et al. [12]. Fig. 3 illustrates a 2D projection of our 3D system where liquid particles are confined between two parallel walls perpendicular to the *z*-direction. Maffioli et al. modelled the thermal fluctuations of the solid particles binding them to massless lattice sites through springs. The system's size in this case is $(L_x, L_y, L_z) = (15.0819, 15.3898, 22.6229)$ in Lennard-Jones units. For the NEMD simulations, a constant velocity in the *x*-direction is applied to the wall lattice sites and a reduced time step of $\delta t = 0.001$ is employed to integrate the equations of motion. A Langevin thermostat with a damping factor of $100 \times \delta t$ is used on the solid wall particles to maintain the temperature of the walls constant at $T = 1.1$ in reduced units. The equations of motion of the fluid are unthermostatted so energy is dissipated through the walls to mimic a realistic system. We found that simulating $3 \times 10^5$ nonequilibrium trajectories allowed a root-mean-square signal-to-noise ratio greater than 10 for the steady-state velocity across the channel, for all the systems we investigated. To perform our simulations, 600 independent equilibrium trajectories were simulated in the NVT ensemble. After an equilibration interval equal to $1 \times 10^6$ timesteps for each of these trajectories, 125 equilibrium states were sampled every $1 \times 10^4$ time steps, from every independent trajectory. Each sampled state was then mapped in 4 different ways, as explained in [12]. Thus 4 independent nonequilibrium trajectories are generated at a starting time $t = 0$ for each of the 125 equilibrium sample states.

The atomic interactions are modelled through a Lennard-Jones potential of the form

$$V(r) = 4\varepsilon \left[ \left( \frac{\sigma}{r} \right)^{12} - c \left( \frac{\sigma}{r} \right)^{6} \right] \tag{20}$$

where $r$ is the particle-particle distance and $c$ is a coefficient modulating the attractive component of the interaction. We refer to $c$ as the cohesion parameter when considering fluid-fluid interactions ($c_{ff}$) and as the wetting parameter when considering fluid-solid interactions ($c_{fs}$), while for solid-solid interactions $c = 1$. The other two parameters were tuned to ensure that a full range of interfacial slip conditions could be modelled, maintaining consistency with the canonical Couette flow for shearing plates. The cohesion parameter is $c_{ff} = 1.3$ for all the simulations, whereas different values were used for the wetting parameter $c_{fs}$, varying between 0.35 and 1.00. We set $\sigma = \varepsilon = 1$ and use a cut-off radius of 2.0. In what follows for the Lennard-Jones system, all quantities are reported in reduced units. The simulation box comprises 648 solid particles and 648 massless lattice sites, while the number of fluid particles varies between 2318 and 2592 for different $c_{fs}$ to guarantee a consistent mid-channel fluid density amongst the systems. Results are obtained for two distinct boundary conditions (BCs) as shown in Fig. 1. In case (b) the wall velocity is twice the value that it is in case (a), to ensure the two different boundary driven systems have the same steady-state shear rate of $\dot{\gamma} = 0.01$. The shear rate was chosen low enough to guarantee a negligible shear heating and to avoid shear thinning. The temperature increase in the centre of the channel was always lower than 0.5% of the wall temperature. For $c_{fs} = 0.35$ and boundary setup as in Fig. 1(b), the number of trajectories was limited to $1 \times 10^5$, due to the length of the simulations.

For the water and graphene system, the same algorithm is employed simulating a total of 1 million trajectories. Here the walls are made of 3 graphene layers. The external layer of each wall is fixed, constraining the volume of the system. The two inner-most layers are thermostatted in the cross-stream directions with a Langevin thermostat, keeping the temperature at $T = 300K$. The Tersoff three-body potential is used to model the intra-layer carbon-carbon interactions [44, 45]. 320 water molecules are confined between the two carbon walls. The SPC/E model [46, 47] is used for the water molecules, constrained with the SHAKE algorithm [48]. The water density is set to 0.95 $g/cm^3$, slightly under the system equilibrium value at atmospheric pressure (0.97 $g/cm^3$). A Lennard-Jones plus long-range Coulombic pair potential with a cut-off of $r_c = 10$Å was employed to describe the interactions between the partial charges of the water molecules and the graphene layers, as well as the interactions between the graphene layers themselves. The Lennard-Jones parameters were also taken from reference [49]. The long-range Coulombic interactions were calculated using the particle-particle particle-mesh (PPPM) algorithm [50]. The walls move at constant velocity $\pm\hat{u}/2 = 50$ m/s and $\hat{u} = 100$ m/s for the counter-sliding walls and single-sliding wall BCs, respectively (see Fig. 1).

All simulations were run with the open source software LAMMPS [51]. The water and graphene system was set up using VMD [52].

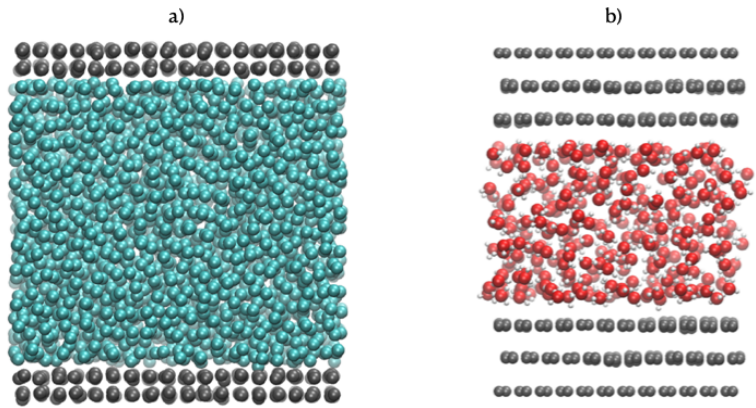

**Figure 3:** (a) Lennard-Jones system for $c_{fs} = 1.00$ and (b) water confined in graphene.

## 3.2 Continuum models and fitting method

We considered three different boundary conditions: no-slip, slip and time-dependent slip BCs. Even for the highest wetting coefficient we adopted ($c_{fs} = 1$), MD results for the Lennard-Jones system showed a certain amount of slip. Consequently, the no-slip model was not suited to describe most of the MD results of this study. The Navier slip model instead describes the amount of slip through the slip length $L_s$. However, $L_s$ is assumed to be constant in time and equal to its steady-state value. We found that this assumption limits the accuracy of the model for time-evolving MD systems, where $L_s$ relaxes over time. We verified in our simulations the relaxation of the slip length, shear stress and friction coefficient to their steady-state values but do not present the results here for reasons of brevity. Given these considerations, to model slip relaxation we applied time-dependent BCs to the Navier-Stokes velocity, described as follows.

In our simulations, two fluid slabs near the walls were designated as boundaries to avoid the interfacial fluid layers where particle velocities and the fluid density are affected by the confinement. As a result, the new boundary conditions are no longer time-independent and we describe them with the following functional form

$$u_{slab}(t) = \bar{u}_{slab}\left[1 - \sum_{k=1}^{N} \alpha_k^{slab} \exp(-\beta_k^{slab} t)\right] \tag{21}$$

where $\bar{u}_{slab}$ denotes the steady state velocity of the fluid slab. Parameters $\alpha_k^{slab}$ and $\beta_k^{slab}$ were fitted to the MD results for each fluid slab. For the counter-sliding walls setup, the fitting parameters resulted from the average of the absolute value of the two fluid slabs' evolution. We found that for $N = 4$, Eq. (21) was able to accurately describe the relaxation of the fluid slab velocities to the steady-state values of every system. Due to confinement effects - which influence the density and velocity profiles near the walls - $\bar{u}_{slab}$ was adjusted based on a linear fit of the velocity profile taken at the center of the channel. Using Eq. (21) for the boundary conditions, Eq. (17) is restated as follows:

$$\begin{cases} \dfrac{\partial u}{\partial t} = \nu \dfrac{\partial^2 u}{\partial z^2} & z \in [-L/2, L/2] \text{ and } t > 0, \\ u(-L/2, t) = u_B(t) & t > 0, \\ u(L/2, t) = u_T(t) & t > 0, \\ u(z, 0) = 0 & z \in [-L/2, L/2], \end{cases} \tag{22}$$

where it is noted that $u_B(t)$ and $u_T(t)$ are the bottom and top slab velocities described by Eq. (21), respectively. Eq (22) leads to the following analytic solution

$$\begin{aligned} u(z,t) = {} & \frac{u_T(t) + u_B(t)}{2} + \frac{u_T(t) - u_B(t)}{L} z + \frac{2}{L}\Bigg[\sum_{n=0}^{\infty} \frac{(-1)^{n+1}}{\lambda_{s,n}^{(0)}} \sum_{k} \left(\bar{u}_T \Gamma_k^{s,L} \Lambda_k^{s,L} + \bar{u}_B \Gamma_k^{s,B} \Lambda_k^{s,B}\right) \cos(\lambda_{s,n}^{(0)} z) \\ & + \sum_{n=0}^{\infty} \frac{(-1)^{n+1}}{\lambda_{a,n}^{(0)}} \sum_{k} \left(\bar{u}_T \Gamma_k^{a,T} \Lambda_k^{a,T} - \bar{u}_B \Gamma_k^{a,B} \Lambda_k^{a,B}\right) \sin(\lambda_{a,n}^{(0)} z)\Bigg], \\ \text{with } & \Gamma_k^{j,i} = \frac{\alpha_k^i \beta_k^i}{\nu(\lambda_{j,n}^{(0)})^2 - \beta_k^i} \text{ and } \Lambda_k^{j,i} = \exp(-\beta_k^i t) - \exp[-\nu(\lambda_{j,n}^{(0)})^2 t] \text{ for } i \in \{B, T\}, \ j \in \{s, a\}. \end{aligned} \tag{23}$$

Eq. (23) was fitted to the MD results, using the viscosity as a free parameter. This choice was made because of the well known nonlocal viscosity inhomogeneity in confined MD systems that takes place especially near the walls, where the fluid density oscillates the most [53–55]. It is important to stress that the viscosity we use here is essentially an effective (constant) viscosity that does not account for the intrinsic nonlocality of the true viscosity kernel [56–60]. Uncertainty propagation was evaluated by block averaging the set of trajectories, producing 50 block averages that were used to perform the fitting independently. The bootstrap method was then applied to assess the distribution of the viscosity and other quantities of interest.

# 4 Results

## 4.1 Lennard-Jones Fluid

We simulated five different systems, varying the wetting coefficient as explained in Section 3.1, and using the two different BC setups, shown in Fig. 1. The relevant steady-state velocity profiles are presented in Fig. 4. Adjacent to the walls we highlight the fluid region where the velocity is affected by the confinement, diverging from the linear continuum prediction. We disregarded all data in this region for the purposes of quantitative analysis.

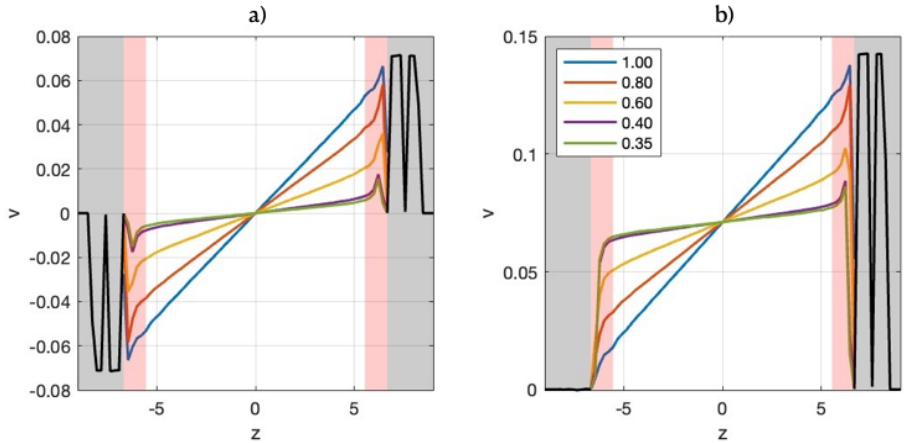

**Figure 4:** Steady-state velocity profiles for a) counter-sliding walls setup and b) single-sliding wall setup, for different wetting coefficient $c_{fs}$, depicted by different coloured curves. The grey shaded regions highlight the wall layers and the red regions the fluid slabs disregarded for the fitting.

Each system exhibits a different amount of interfacial slip that is quantified by the slip ratio, defined here as the ratio of the slip velocity and the wall velocity ($\hat{u}$), both measured relative to the velocity of the fluid's centre of mass:

$$R_s = \frac{\hat{u} - \lim_{t \to \infty} u(z,t)|_{wall}}{\hat{u} - u_{cm}}, \qquad (24)$$

where $u_{cm}$ is the velocity of the fluid's centre of mass and $u(z,t \to \infty)|_{wall}$ is extrapolated from the linear fit of the steady-state velocity profile projected onto the interface. The interface is defined as the location at which the wall density approaches zero. When a linear steady-state velocity profile is assumed, the slip ratio relates to the slip length ($L_s$) as

$$R_s = \frac{2L_s}{2L_s + L}. \tag{25}$$

The advantage of introducing this new quantity lies in the ease with which one can quantify the amount of slip and compare it across different systems, without the need for further nondimensionalisation. The slip ratio characterising every system is reported in Table 1. Lower fluid-solid interactions correspond to higher slip ratios. In the limiting cases, $R_s = 0$ for purely no-slip interfaces, while $R_s = 1$ for a system that displays infinite slip.

| $c_{fs}$ | 1.00 | 0.80 | 0.60 | 0.40 | 0.35 |
|---|---|---|---|---|---|
| $R_s$ | 0.13 | 0.38 | 0.67 | 0.88 | 0.91 |

**Table 1:** Slip ratio for every wetting coefficient value for the Lennard-Jones system. The values reported are the average of the values computed for the two setups. The 95% confidence intervals are smaller than 2.5% of the mean value and not reported here.

In all the results that follow, we fit our NEMD velocity profiles only to the analytical solution for the time-dependent BCs, i.e. Eq (23), because the no-slip and constant-slip solutions proved to be inadequate to obtain accurate fits. The fluid slabs used to fit Eq. (21) were chosen to avoid the regions where confinement effects are evident. Nonlinear least-squares fittings were performed using MATLAB [61]. The continuum model in Section 3.2 was then used to fit the NEMD data as is shown in Fig. 5 for $c_{fs} = 1.00$. Eq. (23) was also used to evaluate the time needed by each system to reach the steady-state. The saturation time of a system is computed as the amount of time required for the spatial integral of the absolute value of the fluid velocity to reach the 99% of its steady-state value. The time-evolution of the velocity and its saturation time for $c_{fs} = 1.0$ are reported in Fig. 6.

Good fits between Eq. (23) and the other systems' results were also found, though we do not present all profiles to keep the results as succinct as possible. The effective viscosity value, resulting from the fitting, decreases with lower wetting coefficient (Fig. 7). The reason for this is found in the lower fluid-solid interaction, causing a lower fluid density near the walls, which translates into a lower overall effective viscosity and a loss of momentum transfer that affects the time evolution of the velocity profile, as represented in Figs. 5 and 6.

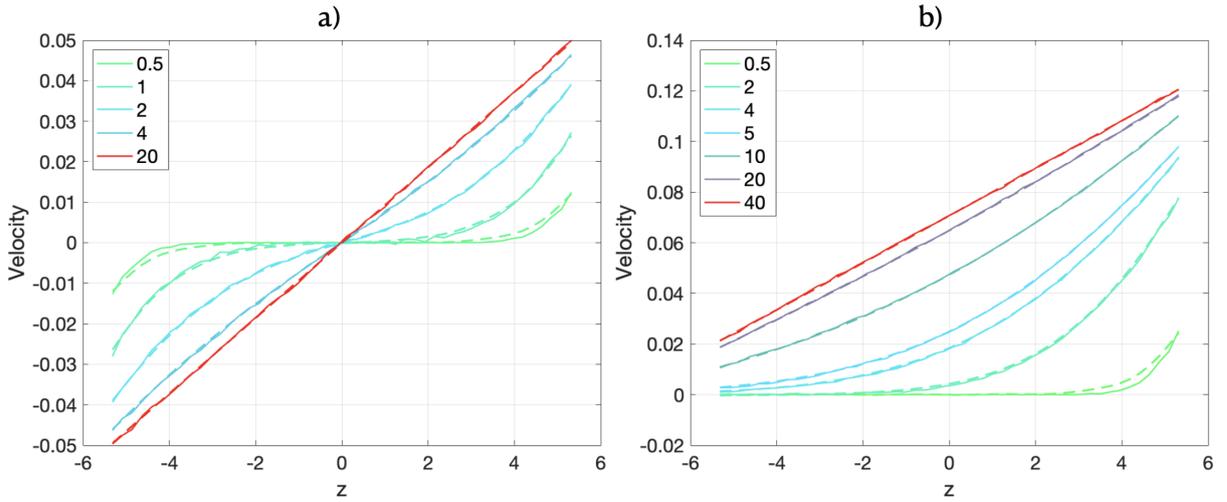

**Figure 5:** Velocity profiles at different times in Lennard-Jones units for $c_{fs} = 1.00$. Continuous lines are used for NEMD data and dashed lines for the analytical continuum model, Eq (23). (a) counter-sliding sliding walls; (b) single sliding wall. The different evolution times are depicted in the legend insets by different colours.

To characterize the effect of boundary slip on the dynamical response, we consider the saturation time ratio between the counter-sliding and single-sliding wall setups. This ratio exhibits an asymptotic behaviour as $R_s \to 1$, as shown in Fig. 8. Accordingly, for $c_{fs} = 1.0$ and a slip ratio of $R_s = 0.13$, the saturation time ratio reaches a value of 4.04, converging to the no-slip reference value of 4 given by Eq. (8).

### 4.2 Water-Graphene System

For the system of water confined by graphene sheets, we studied the same boundary setups, focusing on the transient evolution of the system. In a fully developed steady-state, water exhibits a slip ratio of $R_s = 0.982$ on graphene sheets, which highlights its strongly hydrophobic nature. We observe that the two boundary-driven systems go through two different types of transient behaviour. While the velocity evolution of the single sliding wall system resembles that observed in the Lennard-Jones system (see Figs. 9 and 10 (b)) the counter-sliding walls system does not (see Fig. 10 (a)). The overshoot for the counter-sliding walls system means that the same techniques based on the assumed functional forms of the transient solutions of the Navier-Stokes equations used to estimate the viscosity of the Lennard-Jones system cannot be used for the water-graphene system. This meant that no good fit of the data to Eq. (21) could be made. Instead, the saturation time of the system was directly evaluated from the

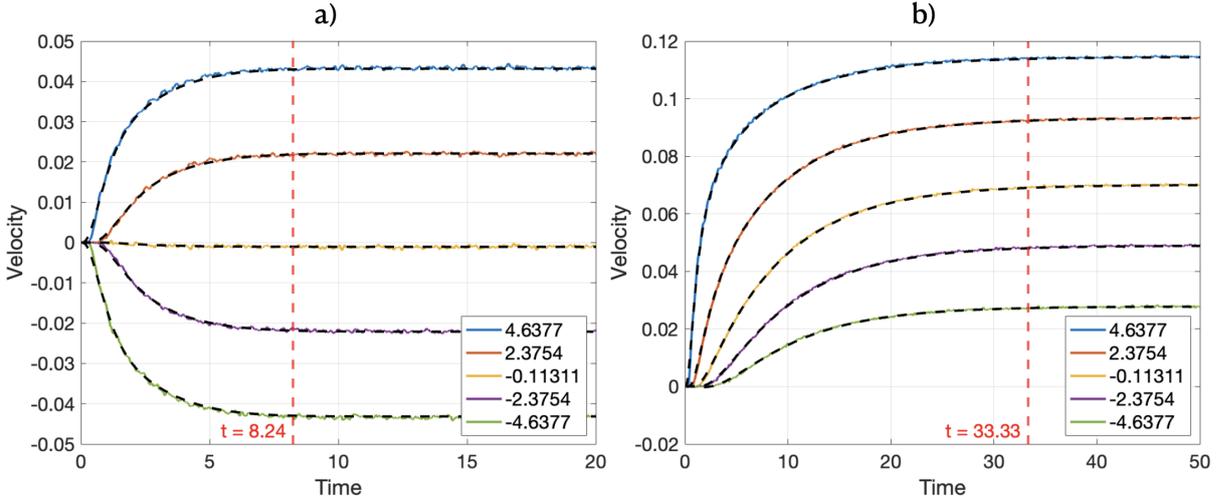

**Figure 6:** Velocity evolution at different channel coordinates (in reduced Lennard-Jones units), depicted by different colours as shown in the legend inserts. The dashed red vertical line indicates the saturation time for the 99% threshold. Continuum lines are for NEMD data, while dark dashed lines represent the continuum model predictions. (a) Counter-sliding walls; (b) single-sliding wall.

MD data, by considering the spatial integral of the absolute value of the velocity (Fig. 11). The single sliding wall and the counter-sliding walls setups have shown, respectively, an average saturation time of 0.370 ns and 0.003 ns. Note that the saturation time for the counter-sliding wall setup is intentionally overestimated to ensure a conservative estimate. As a result, the saturation time ratio may be slightly underestimated. The saturation time ratio is reported in Fig. 12 and compared with the previous results for the Lennard-Jones system. As anticipated in the continuum model in Section 2, the wave number of the symmetric component converges to 0 as the degree of slip increases. For this reason, the saturation time ratio exhibits an asymptotic behaviour for slip ratios close to 1. Despite the limited amount of data, we performed a weighted least-squares fit on the saturation time ratios obtained from the Lennard-Jones systems, using an asymptotic empirical function of the form:

$$S_r(R_s) = \frac{A}{(1-R_s)^b} + B, \qquad (26)$$

subject to the constraint $A + B = 4$, which enforces the no-slip reference condition. Here $A = 1.4508$, $B = 2.5492$ and $b = 1.1545$. Fig. 12 shows that the saturation-time ratio for water on graphene falls, to a good approximation, within the prediction of the fitted model. It demonstrates that a system of water confined to graphene walls driven by walls moving with equal and opposite velocities will reach steady-state about 120 times faster than an equivalent system driven by only one moving wall. The universality of Eq. (26) suggests that systems with high degrees of slip will always reach steady-state orders of magnitude faster

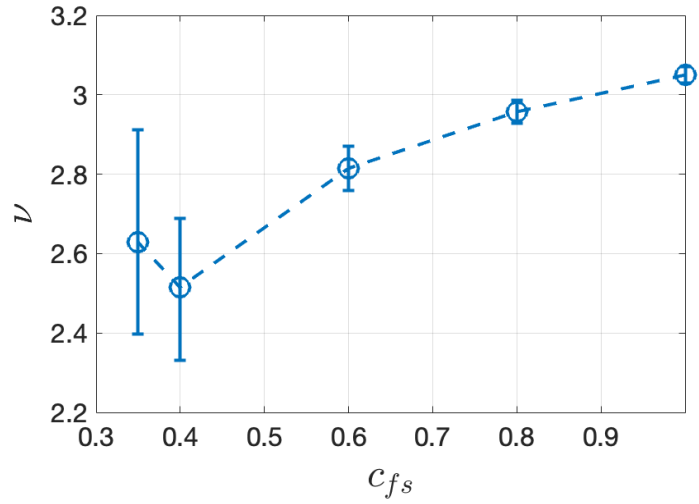

**Figure 7:** Effective viscosity found from the fit of Eq. (23) to the MD data with 95% confidence intervals. The values shown here are the result of the average of the effective viscosity found for the two boundary setups.

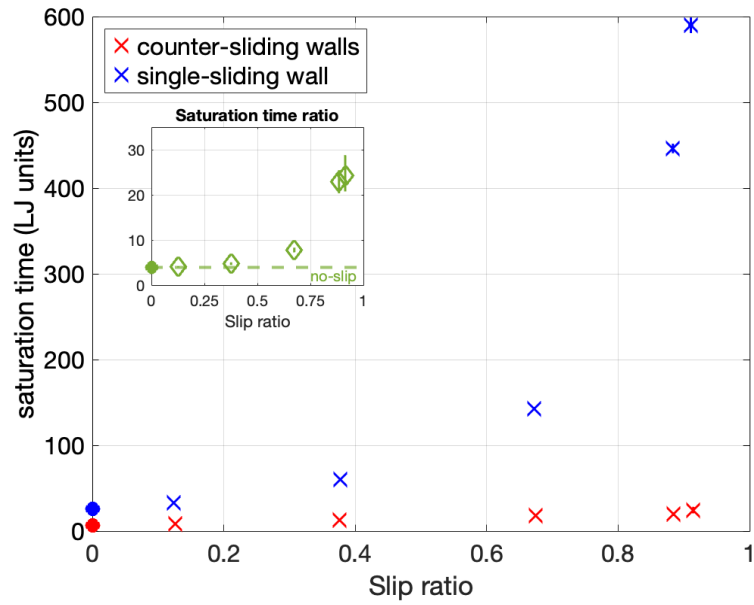

**Figure 8:** Saturation time vs slip ratio for the two boundary driven setups for the Lennard-Jones system. The inset displays the saturation time ratio between the counter-sliding and single-sliding wall systems. Filled circles represent the values obtained from the no-slip model using $\nu = 3.05$. The 95% confidence intervals are shown; when not visible, they are smaller than the symbol size.

by shearing both walls compared to just one.

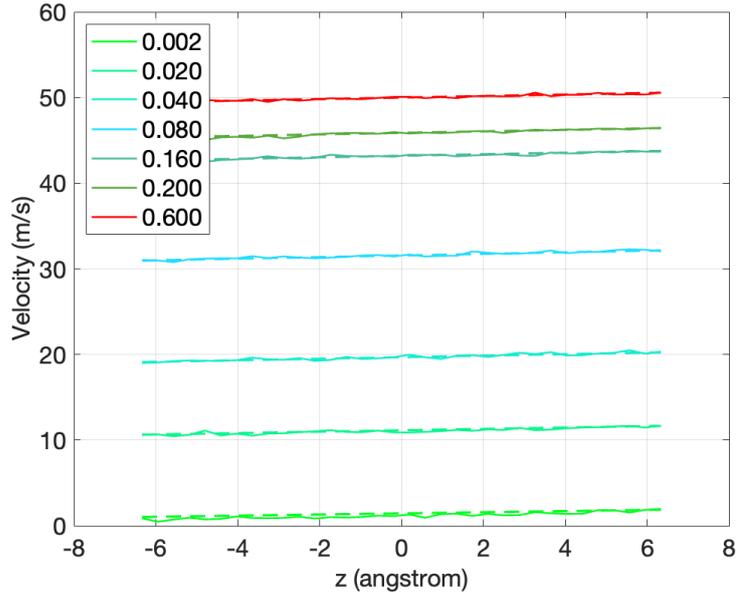

**Figure 9:** Velocity profile at different times (ns) for water on graphene when the single sliding wall setup is employed.

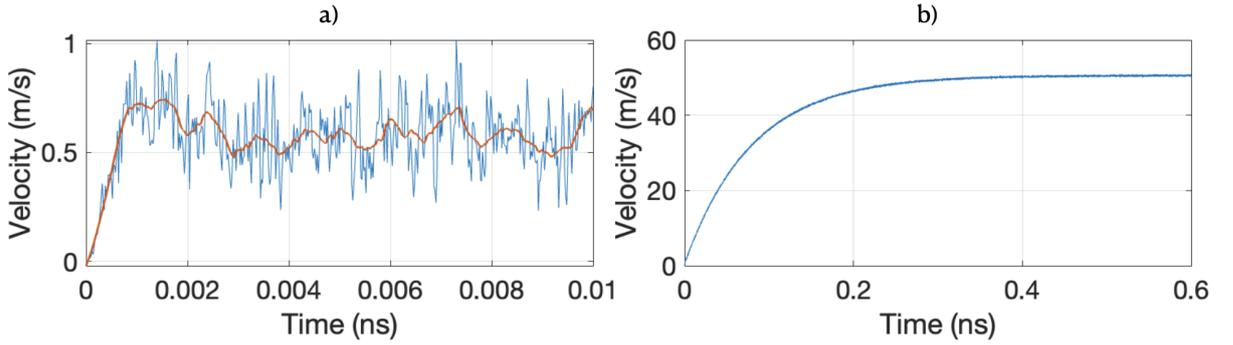

**Figure 10:** Velocity evolution at $z = 6.34$ Å for the system of water on graphene. Results for the (a) counter-sliding walls and (b) single-sliding wall setups are shown. The red line in (a) shows the filtered signal obtained by using a Savitzky–Golay filter [62, 63].

## 5 Conclusion

In this study we examined how the symmetry of boundary conditions affects the time it takes a boundary-driven planar Couette flow system to reach its steady state, and how this effect is amplified when interfacial slip is considered. To this end, we investigated different continuum models, supported by nonequilibrium molecular dynamics simulations of a Lennard-Jones system and a widely studied realistic system, namely water confined to graphene walls. We explored the full range of slip and validated the continuum predictions using our NEMD simulations. Our results demonstrate that the counter-sliding walls setup is significantly more efficient than a single-sliding wall configuration when simulating boundary-driven sheared systems. In fact, a system with counter-sliding walls will always reach steady state in finite time, regardless of the amount of slip and despite the associated loss of momentum transfer. This behaviour is predicted by continuum models and is fully confirmed by our MD results. The equilibration time gain is independent of the molecular details of the fluid and is hydrodynamic in nature. It is fourfold for systems near the no-slip regime and increases dramatically for systems exhibiting high slip. For a water-graphene system the speed-up is over two orders of magnitude, but in principle it could be much more and can be determined from a universal functional form given by Eq. (26). For such systems — especially when the steady state is of primary interest — choosing an appropriate boundary condition symmetry allows computational resources to be redirected toward simulating a broader ensemble and improving statistical accuracy. Moreover, response theory techniques such as the transient-time correlation function (TTCF) method, for which the signal-to-noise ratio decreases with the length of the simulated time interval, can retain their remarkable accuracy by focusing on systems with shorter time evolutions to attain steady-state. This is especially relevant when simulating high molecular weight systems, such as lubricants sheared at experimentally accessible strain rates, or for high-slip systems such as water confined to graphene channels or carbon nanotubes, where the transient response time can be substantially reduced by choosing counter-sliding walls. Since shear flow is the diffusive transport of transverse momentum the principles discussed here can equally be applied to other diffusive transport phenomena. One example of this is heat flow, where we could expect the establishment of steady-state heat flow to be faster for a given temperature difference across a system of fixed width with a symmetric, rather than asymmetric temperature difference. This can be achieved by simultaneously decreasing the temperature at one wall by $\Delta T/2$ and increasing the other by $\Delta T/2$ rather than increasing only one of them by $\Delta T$.

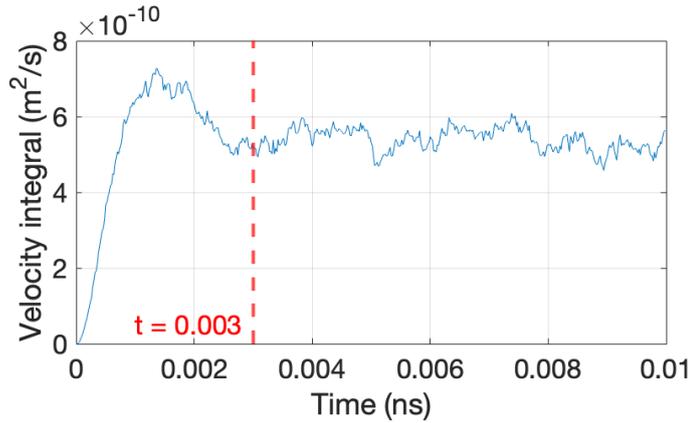

**Figure 11:** Spatial integral of the absolute value of velocity for the water-graphene system when the counter-sliding walls setup is employed. The red dashed line indicates the chosen saturation time.

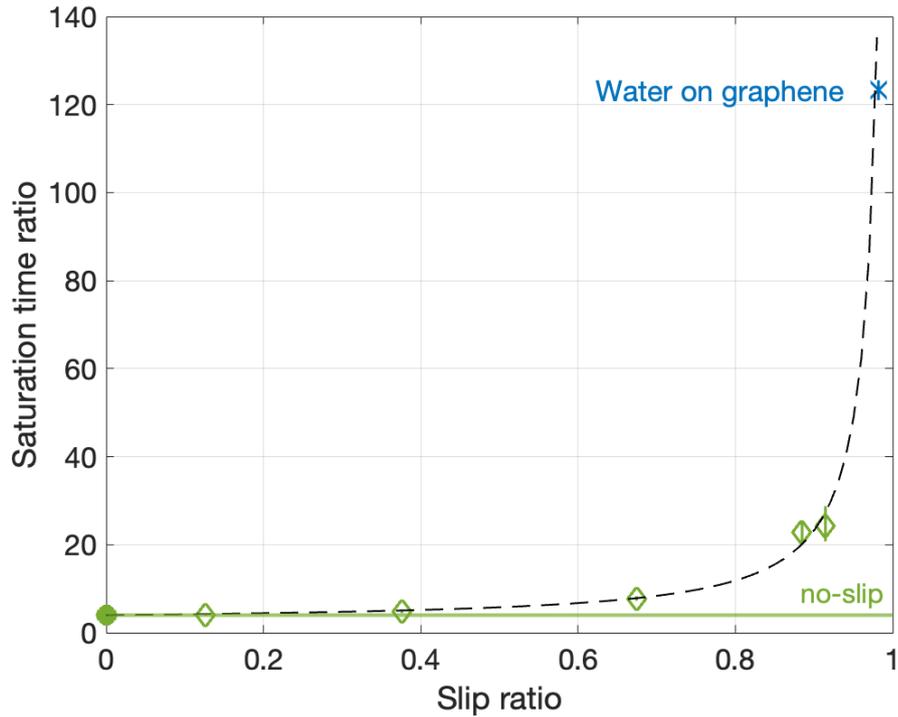

**Figure 12:** Saturation time ratio for the different systems investigated. Filled circle for no-slip model where $\nu = 3.05$ was used, empty diamonds for Lennard-Jones system results fitted with the time-dependent BCs, and the "x" for the water-graphene system. The 95% confidence intervals are shown; when not visible, they fall below the marker size. The black dashed line represents Eq. 26 fitted to the Lennard-Jones data.

Again, we expect that the effect would be even more pronounced in the presence of a temperature jump (Kapitza resistance) at the walls. Finally, we note that the observations made in this study are not just of theoretical interest. For example, there are potential impacts in applications such as roll-on film coating, where a rolling top would be less efficient than having both surfaces (i.e., the coated surfaces themselves) spin; ink jets and painting where changing speeds in counter-spinning rollers risk blockages which can be reduced by controlling slip; and in tribometers, where measurements cannot be taken in the transient period.

## Supplementary Material

Supplementary material containing full derivations of the underlying hydrodynamic solutions to the governing differential equations discussed in Sections 2 and 3 is in the accompanying pdf file "Supplementary Material - Theory". Further information containing tables of data resulting from our simulations and analysis is contained in the file "Supplementary Material - Data".

## Acknowledgements

We thank the Australian Research Council for a grant obtained through the Discovery Projects Scheme (Grant No. DP200100422) and the Royal Society for support via International Exchanges, Grant No. IES/R3/170/233. J.P.E. was supported by the Royal Academy of Engineering (RAEng) through their Research Fellowships scheme. D.D. was supported through a Shell/RAEng Research Chair in Complex Engineering Interfaces. The authors acknowledge the Swinburne OzSTAR Supercomputing facility and the Imperial College London Research Computing Service (DOI:10.14469/hpc/223) for providing computational resources for this work.

## Data Availability

Data supporting the findings of this study is available in the accompanying file "Supplementary Material - Data". Further information is available from the corresponding author upon reasonable request.

## Appendix A: Saturation time ratio law - first order analytic approximation

We can derive an approximate expression for our empirical equation, Eq. (26), from first principles. Equation (15), for the zero wave numbers is recalled here as:

$$\tan\left(\tilde{\lambda}_{s,0}\right) = \frac{1}{\tilde{L}_s \tilde{\lambda}_{s,0}} \quad \text{and} \quad \cot\left(\tilde{\lambda}_{a,0}\right) = -\frac{1}{\tilde{L}_s \tilde{\lambda}_{a,0}},$$
$$\text{for } \tilde{\lambda}_{s,0} = \frac{\lambda_{s,0} L}{2}, \quad \tilde{\lambda}_{a,0} = \frac{\lambda_{a,0} L}{2} \quad \text{and} \quad \tilde{L}_s = \frac{2 L_s}{L}, \tag{27}$$

from which, it follows that

$$\begin{cases} \lim_{\tilde{L}_s \to \infty} \tilde{\lambda}_{s,0} = 0, \\ \lim_{\tilde{L}_s \to \infty} \tilde{\lambda}_{a,0} = \frac{\pi}{2}. \end{cases} \tag{28}$$

By taking a Taylor series expansion, we have

$$\tan\left(\tilde{\lambda}_{s,0}\right) = \tilde{\lambda}_{s,0} + \ldots \qquad \text{for } \tilde{\lambda}_{s,0} \to 0,$$
$$\cot\left(\tilde{\lambda}_{a,0}\right) = -\left(\tilde{\lambda}_{a,0} - \frac{\pi}{2}\right) + \ldots \quad \text{for } \tilde{\lambda}_{a,0} \to \frac{\pi}{2}. \tag{29}$$

Hence, disregarding all higher orders terms, we can re-write Eq. (27) as

$$\tilde{\lambda}_{s,0} = \frac{1}{\tilde{L}_s \tilde{\lambda}_{s,0}} \text{ and } -\left(\tilde{\lambda}_{a,0} - \frac{\pi}{2}\right) = -\frac{1}{\tilde{L}_s \tilde{\lambda}_{a,0}}. \tag{30}$$

Since we are considering $\tilde{L}_s \to \infty$, as stated in Eq. (28), the second equation in Eq. (30) simplifies to $\tilde{\lambda}_{a,0} = \frac{\pi}{2}$. Hence we obtain, to first order in the expansion,

$$\tilde{\lambda}_{s,0}^2 = \frac{1}{\tilde{L}_s} \text{ and } \tilde{\lambda}_{a,0} = \frac{\pi}{2}. \tag{31}$$

By evaluating the ratio of the characteristic times of the slowest exponentials we get

$$\frac{\tilde{\lambda}_{a,0}^2}{\tilde{\lambda}_{s,0}^2} = \frac{\pi^2 \tilde{L}_s}{4} = \frac{\pi^2}{4} \frac{R_s}{(1-R_s)}, \tag{32}$$

where in the last equality we make use of Eq. (25). Fig. (13) shows how Eq. (32) compares with the empirical formula given by Eq. (26). More accurate (albeit more complicated) approximations could of course be made by taking higher order Taylor expansions or by a perturbative expansion, but we do not do this here. Despite the simplified first-order Taylor expansion presented here, Fig. 13 demonstrates it is a reasonable approximation to represent the observed results, particularly at higher slip lengths.

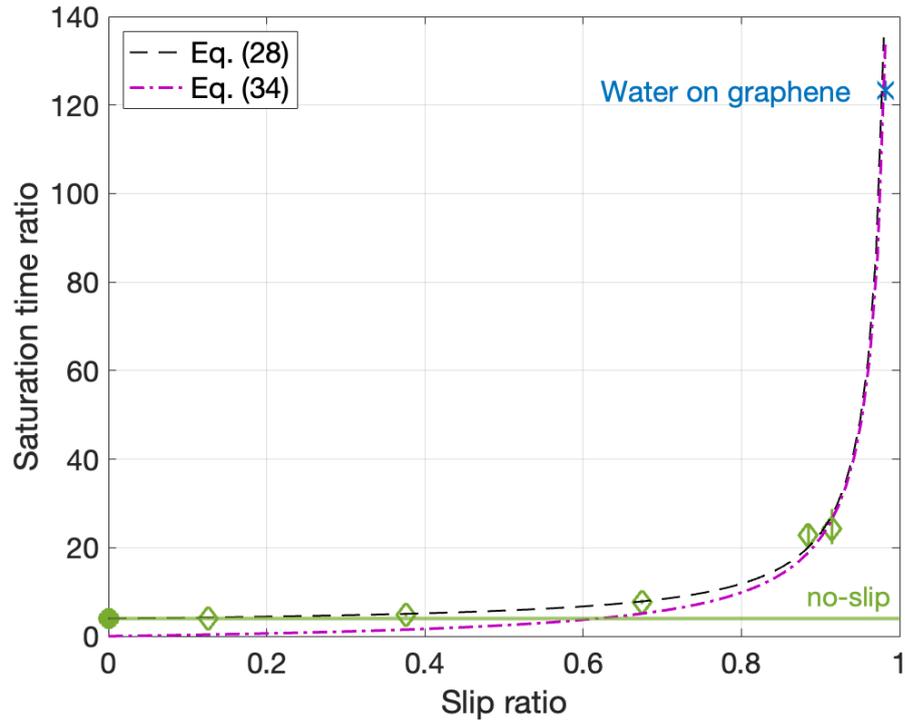

**Figure 13:** Saturation time ratio for the different systems investigated. Filled circle for the no-slip model where $\nu = 3.05$ was used, empty diamonds for Lennard-Jones system results fitted with the time-dependent BCs, and the "x" for the water-graphene system. The 95% confidence intervals are shown; when not visible, they fall below the marker size. The black dashed line represents Eq. (26) fitted to the Lennard-Jones data. The purple dash-dotted line represents Eq. (32).